\documentclass[epj,nopacs]{svjour}
\usepackage{graphicx,amsmath}
\usepackage{wasysym,pifont}
\usepackage{epsfig}
\usepackage{newlfont,bm,dcolumn,longtable}
\usepackage{amssymb,amsfonts,amsmath}
\usepackage{color}
\usepackage{eucal}

\newcommand{\be}{\begin{equation}}
\newcommand{\ee}{\end{equation}}
\newcommand{\bea}{\begin{eqnarray}}
\newcommand{\eea}{\end{eqnarray}}

\newcommand{\thav}[1]{\left< #1 \right>}

\begin{document}

\title{
 A numerical study of planar arrays of correlated spin islands
}

\author{I. Maccari\inst{1} \and A. Maiorano\inst{1,2} \and E. Marinari\inst{3} \and{J. J. Ruiz-Lorenzo}\inst{4,2}}

\institute{Dipartimento di Fisica,
  Sapienza Universit\`a di Roma, P. A. Moro 2, 00185 Roma, Italy \and
Instituto de Biocomputaci\'on y F\'{\i}sica de Sistemas Complejos (BIFI)
  50018 Zaragoza, Spain \and
Dipartimento di Fisica, IPCF-CNR and INFN,
  Sapienza Universit\`a di Roma, P. A. Moro 2, 00185 Roma, Italy \and
Departamento de F\'{\i}sica and 
Instituto de Computaci\'on Cient\'{\i}fica Avanzada (ICCAEx),
 Universidad de Extremadura, 06071 Badajoz, Spain
}

\date{\today}

\abstract{
We analyze a system of interacting islands of XY spins on a triangular
lattice. This model has been introduced a few years ago by Eley et
al. to account for the phenomenology in experiments on tunable arrays
of proximity coupled long superconductor-normal metal-superconductor
junctions. The main features of the model are the separation of a
local and a global interaction energy scale and the mesoscopic
character of the spin islands. Upon lowering the temperature the model
undergoes two crossovers corresponding to an increasing phase
coherence on a single island and to the onset of global coherence
across the array; the latter is a thermodynamical phase transition in
the Ising universality class.  The dependence of the second transition
on the island edge-to-edge spacing is related to the proximity-effect
of the coupling constant.
}

\maketitle

\section{Introduction}
Recently Eley et al.~\cite{ELEY1} have introduced a model of coupled
islands of XY spins. Their goal was explaining the results of
measurements of resistance in arrays of long
superconducting-normal-superconducting junctions.  The experimental
devices are based on planar arrays of identical islands made of
superconducting (Nb) grains, disposed in a triangular matrix over a
metal (Au) film. The authors studied the dependence of the system
resistance $R(T;h,\ell)$ on temperature $T$, on island (vertical)
thickness $h$ and on inter-island spacing $\ell$.  They found i) the
resistance dropping to zero, by lowering $T$, in two steps, and they
determined two transition temperatures $T_1$ and $T_2$ with $T_1>T_2$;
ii) an interesting dependence of both $T_1$ and $T_2$ on the island
spacing, possibly (both) extrapolating to $T=0$ at large $\ell$; iii)
a strong dependence of the behavior of the system on the island
thickness. In a following paper~\cite{ELEY2_PUB}, they discussed a
more detailed comparison between the experimental data and the
predictions about the dependence of $T_2$ on $\ell$ given by the
conventional theory of Lobb, Abraham, and Tinkham
(LAT)~\cite{LOBB}. They argued that for large inter-island spacing the
superconducting transition is more likely to be driven by diffusive
effects~\cite{WILHELM,DUBOS} in the normal metal substrate, and that
it does not depend on the details of the superconducting islands, with
the puzzling dependence on island height as a notable exception.

The superconducting transition in proximity-coupled \emph{macroscopic}
grains embedded in normal metal films has been the object of intensive
work in the past years,~\cite{SPIVAK1,SPIVAK2,FEIGELMAN}. Tunable
realizations of $2D$ superconductivity were also object of previous
experiments~\cite{RESNICK}. The classical model presented in
Ref.~\cite{ELEY1} to account for a novel phenomenology is at
difference with previous theoretical and experimental work, as it
takes into account the intrinsic fluctuations of the superconducting
state inside the single \emph{mesoscopic} islands (see also
refs.~\cite{GRAPHENE1,GRAPHENE2} for recent theoretical and experimental work
on mesoscopic Sn islands laid on graphene). 
It is clear that, because of many reasons we will discuss in the
following, this model does not try to reproduce faithfully the
experimental situation (for example the use of an anisotropic coupling
is not connected to the physical form of the Josephson interaction but
is a tool needed to obtain a phase transition).
The idea of \cite{ELEY1}, and our point of
view here, is to analyze a very simple model that offers a behavior
quite similar to the one detected in the experiments, and to try to
learn from this behavior. Here we will present a detailed analysis of
the model, that corroborates and supplements the hints coming from
the first analysis of  \cite{ELEY1}.
It is also worth mentioning that tunable two-dimensional superconductors
are also of interest in a revived search for a non-conventional $2D$,
$T=0$ metallic
phase.~\cite{PHILLIPS,PUNNOOSE,KRAVCHENKO1,KRAVCHENKO2,KRAVCHENKO3,GRILLI_CAPRARA,WEILIU,HAN} 

\section{The model}
The Hamiltonian of the model is based on $O(2)$ vectors living on
the individual grains (labeled with $i,j$). Groups of grains form
islands (labeled by $p$):
\begin{eqnarray}
  \label{eq:H}
  H = & - & 
  J\sum_p\sum_{\langle ij\rangle \in p} \vec{S}_i\cdot\vec{S}_j
  \nonumber  \\ 
  & - & \sum_{\langle p, p^\prime \rangle}
  \vec{M}_p\cdot\tens{J}^\prime \vec{M}_{p^\prime} \; , \\
  \vec{M}_p & \equiv &\sum_{i \in p} \vec{S}_i \; ,
\end{eqnarray}
where by a dot we denote the scalar product in the internal space and
where $\tens{J}^\prime$ is a $2\times 2$ matrix of couplings. Each
island is a $D$-dimensional hyper-cubic array of grains of linear size
$I$ (and volume $V_I=I^D$, with either $D=1$ or $D=2$ in our
computations).  Islands are arranged on a (two dimensional) planar
regular lattice of linear size $L$. Islands are mesoscopic: their
linear size $I$ is not larger than a few grains. Because of that they
may have large global phase fluctuations. The size of the underlying
planar array is macroscopic, $L\gg I$. The case of one-dimensional
island is an exercise useful to understand better the role of island
dimensionality, and does not try to be a description of the
experimental situation. On the contrary the case of two dimensional
islands is probably closer to the experimental situation, where
islands have many layers, but only one or few conjure to build the
inter-island interaction.

The first term of the Hamiltonian is a sum of nearest-neighbor
interactions between grains contained in the same island.  The second
term couples neighboring islands in the array. Each spin in a given
island interacts directly with its neighboring spins in the same
island and with the average spin field of surrounding islands.  In the
model proposed in Ref.~\cite{ELEY1} the inter-island coupling matrix
(in the internal $O(2)$ space) $\tens{J}^\prime$ is anisotropic:
\begin{equation}
  \tens{J}^\prime = 
  \left( 
  \begin{array}{cc}
    J^\prime & 0 \\
    0 & 0 \end{array}
  \right) \; .
  \label{eq:jprime}
\end{equation}
This particular choice polarizes the islands in one specific direction
in internal vector space, changing the nature of the inter-island
phase transition. This is a technically useful choice (since it
carries a phase transition in the game), but it does not aim at
reproducing the details of the physical Josephson interaction.
Finally, notice that in the isotropic case and \emph{if the energy
  scales $J$ and $J^\prime$ are far apart, i.e. $J^\prime\ll J$}, so
that at low temperatures all (mesoscopic) islands are magnetized, we
recover a Kosterlitz-Thouless~\cite{KT} phase transition.

The island-island couplings depend on the temperature and on the
inter-island edge-to-edge spacing, according to the theory of diffusion
of electron pairs in SC-Normal-SC junction. As in the work of
\cite{ELEY1} we take a ``quasi-\emph{proximity-effect}''
\cite{ELEY1,LOBB} form for both couplings; in a proximity interaction
$J^\prime$ would depend on the inverse square of island spacing when
the latter is small, but following \cite{ELEY1} and for the same sake
of simplicity we omit this part of the interaction, that is not
expected to change the nature of the phase transitions here.  We
assume the proximity-effect form also for the grain-grain coupling $J$
and we take the grain-grain distance as the length (lattice) unit (and denote
the inter-island spacing as $\ell$).
\begin{eqnarray}
  J & = & J_0 \exp\left(-\sqrt{T}\right) \; ,\\
  J^\prime & = & J^\prime_0 \exp\left(-\ell\sqrt{T}\right)\ \  .
\label{eq:proximity}
\end{eqnarray}
The choice of an interaction of a proximity-like form implies that
physically grains of the islands are also immersed into a metallic
matrix.  The authors of Ref.~\cite{ELEY1} introduced
the model defined in (\ref{eq:H}) to explain the presence of two
transitions (intra-island, $T_1$, and inter-island coherence, $T_2$)
and the depression of $T_1$ for increasing island spacing $\ell$. Such
a dependence of $T_1$ on $\ell$ has been observed and reported for the
first time in~\cite{ELEY1} (for example in previous experiments on
lead disks on a thin substrate~\cite{RESNICK} where islands were not
mesoscopic, the effect was not observed).  The energy scales $J_0$ and
$J_0^\prime$ must be well-separated: we adjust them in order to
clearly split the high-$T$ and the low-$T$ transition.  We fix $J_0=1$
and vary $J_0^\prime$; in order to easily compare data for different
island sizes, we also take $J_0^\prime=j_0^\prime/V_I$ and adjust the
parameter $j_0^\prime$.

In Ref.~\cite{ELEY1} the authors also give some predictions by analyzing
a $D=1$ islands model, where it turns out that:
\begin{itemize}
\item $T_2 \leq T_1$ provided $J \gg J^\prime$ and islands are small;
\item $T_1 \rightarrow 0$ when $J^\prime \rightarrow 0$. 
\end{itemize}
The second statement is rather counter-intuitive, as it implies that
islands are not superconducting when they are isolated. This implies
that an array of superconducting islands can be superconducting even
if the inter-island spacing is larger than the superconductor
coherence length, but an isolated island of superconducting grains,
where grains are packed closer than islands are in the array, looses
phase coherence. In this respect, the one-dimensional and mesoscopic
character of the islands plays a role, since for macroscopic chains we
must expect $T_1\sim 0$, and, as we will see in the following, for
large $I$ the intra-island coherence is driven by inter-island
ordering (also see the discussion in the Conclusions section).

Another striking aspect of the phenomenology of the
system\cite{ELEY1,ELEY2_PUB} is the dependence of its behavior from
the height of columnar grains. Realistic islands extend in more than
one dimension.  We have analyzed by numerical simulations the behavior
of one and two-dimensional islands. The dependence on thickness may
suggest that it would be interesting to go to $D=3$, too (in case of a
very large value of $I$ this should turn the $T_1$ transition to a
true second-order one). If energy scales are adequately separated
(i.e. $J\gg J^\prime$), this should not change the properties of the
$T_2$ (KT) transition, when phases of grains in the same island are
mutually locked. Mesoscopic islands can then have a crossover at $T_1$
from a disordered to an ordered phase; for $D\geq 2$ and large $I$
this crossover is related to a true thermodynamic transition.

\section{Numerical simulations}
We have studied the model defined in (\ref{eq:H}). By following a very
slow annealing protocol, with constant ratios between adjacent
temperature (a logarithmic annealing scale), we have cooled down the
system in order to get a signal for the two transitions. At each
temperature we collected measurements during the evolution of the
Monte Carlo dynamics. Our Monte Carlo step consists of $n_m$ sweeps of
the whole lattice by single-spin moves Metropolis dynamics, followed
by $n_o$ sweeps by over-relaxation~\cite{AMIT}. The choice of $n_m=10$
and $n_o=12$ have shown to be appropriate for most island and array
sizes considered (and an overkill for the smaller sizes), and allowed
an estimate of integrated auto-correlation times not larger than ten
Monte Carlo steps at most temperatures. Although averages always
stabilize quickly after any temperature change, we drop the first half
of the collected measurements at all $T$ values.  The simulated
annealing protocol, together with over-relaxation, is appropriate to
the needs of this problem.  All observables of interest converge very
fast to a plateau at all temperatures.  Although averages always
stabilize quickly after any temperature change, we drop the first half
of the collected measurements at all $T$ values.

\begin{figure}[!b]
\begin{center}
\includegraphics[width=0.95\columnwidth]{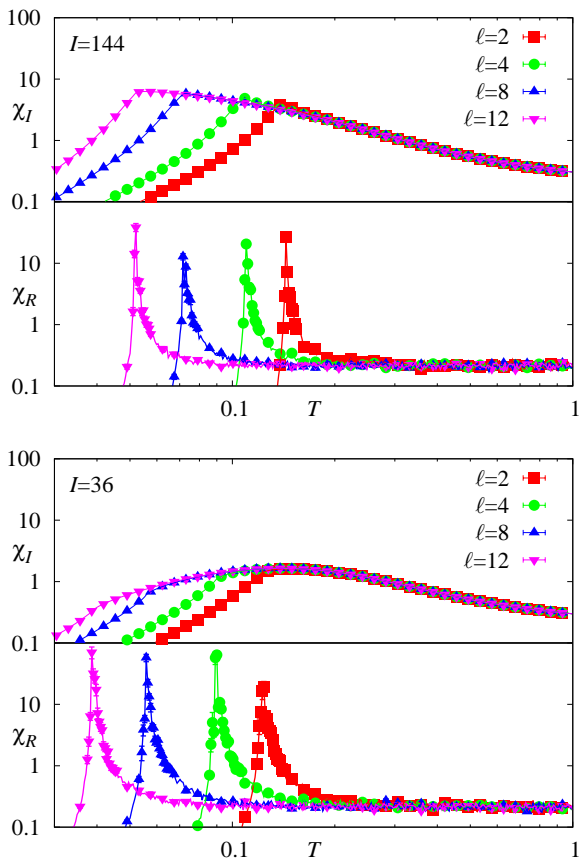}
\caption{$D=1$ results for the largest simulated size $L=32$. We plot
  the connected susceptibilities $\chi_I$ and $\chi_R$ for two
  different values of $I$. $j_0^\prime=0.0072$. On each panel
  we also plot the curves for four values of the island spacing
  $\ell$. Notice that the peaks of $\chi_I$ (which mark the $T_1$
  transition) depend strongly on $\ell$.}
\label{fig:D1}
\end{center}
\end{figure}

Since the devices in the experimental setup~\cite{ELEY1,ELEY2_PUB} are
triangular arrays, we consider a triangular lattice. A simple
implementation choice in simulation is to consider a triangular array
with regular hexagonal shape with helical boundary condition (in this
way we preserve the symmetries of the triangular array and avoid
involved bulk properties extrapolations); each side of the hexagon has
a width of 
$L$ islands, and the number of islands is $V_S=3L(L-1)+1$. We
simulated systems with $L=8,16$ and $32$: for $D=1$ systems we have
islands of sizes $I=16,36,64,100$ and $144$, while for $D=2$
we have $I=6,8,10$ and $12$. The inter-island edge-to-edge spacing $\ell$ has
been varied in the set $\{2,4,8,12\}$ for $D=1$ and
$\{2,4,8,12,16,24\}$ for $D=2$ (the SNS arrays in Ref.~\cite{ELEY1}
had edge-to-edge spacings up to approximately $10$ times the grain size
in their experiments, and $\ell$ up to $20$ in
Ref.~\cite{ELEY2_PUB}). We have considered both free (FBC) and
periodic (PBC) boundary conditions on the single islands. Although we
found no qualitative differences, FBC is a more realistic choice when
dealing with mesoscopic objects, for which we expect finite-size
effects to play a role. We measured the following quantities.
\begin{itemize}

\item The \emph{single island magnetization magnitude} (averaged over islands): 
\be
\label{eq:MI}
M_I=\frac{1}{V_S}\sum_p\left|\frac{1}{V_I}\sum_{i\in p} \vec{S}_i \right|\, .
\ee
This should be, in the limit of infinitely extended islands, a good
order parameter for island internal ordering (in any direction in
internal spin space, and globally over the array: it has a non-zero
value whenever any island starts to order internally and it is maximum
when all islands are locally ordered, independently of the relative
orientation between different islands).

\item The total magnetization:
\be
\label{eq:M}
\vec{M}=\frac{1}{V_SV_I}\sum_p \sum_{i\in p} \vec{S}_i\, .
\ee

\item A renormalized magnetization:
\be
\label{eq:mu}
\vec{\mu}_p = \frac{\sum_{i\in p} \vec{S}_i}{\left| \sum_{i\in p} \vec{S}_i
  \right|}\, ,
\ee
which is a unit vector on the single island, and its average over the array.

\item $\vec{M}_R$, which
  characterizes the globally-ordered phase, even if islands are
not yet internally fully ordered:
\be
\label{eq:MR}
\vec{M}_R=\frac{1}{V_S}\sum_p\vec{\mu}_p\,.
\ee
\end{itemize}

We also consider the fluctuations of the magnetizations
\bea
\label{eq:chi}
\chi & \equiv & V_IV_S\left[\thav{M^2}-\thav{|M|}^2\right] \, ,\\
\label{eq:chiI}
\chi_I & \equiv & V_I\left[\thav{M_I^2}-\thav{M_I}^2\right] \, ,\\ 
\label{eq:chiR}
\chi_R & \equiv & V_S\left[\thav{M_R^2}-\thav{M_R}^2\right] \, ,
\eea
where $\thav{M_I^2}$ and $\thav{M_I}$ are averaged over all
islands. $\chi$ is the total susceptibility of the system. At very low
temperatures, when $M_I\sim 1$, we have $\chi_R \sim \chi/V_I$.  At
$T_2$, that we define as the location of the peak of the inter-island
susceptibility $\chi_R$, $\chi_R$ and $\chi$ have very similar sharp
peaks (both in shape and location).  We take the location of the (very
smooth) maximum of $\chi_I$ as a rough estimate of the temperature
$T_1$ at which islands order internally (in this way we give an
operative definition of $T_1$ in our model: since islands are of
finite extent the $T_1$ defined in this way is indeed a crossover
temperature).

\section{Results and discussion}
In Fig.~\ref{fig:D1} we report the results for arrays of
one-dimensional chains.  The $T_1$ temperature value goes to zero very
fast as the island size grows, as expected for linear spin chains.
Upon lowering $T$, coherence between island builds up and also drives
the internal ordering; the two transitions can be resolved only for
very small island sizes and by lowering considerably the value of coupling
constant $j_o^\prime$. The effect is also strongly dependent on island
size.

\begin{figure}[!t]
\begin{center}
\includegraphics[width=0.95\columnwidth]{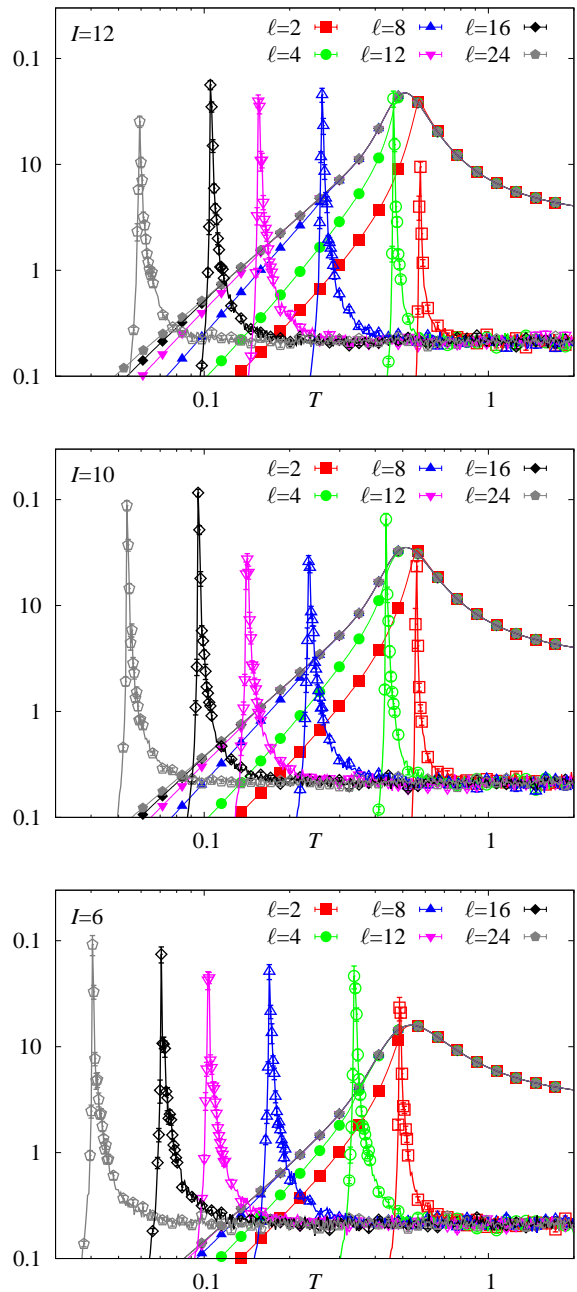}
\caption{$D=2$  results for the largest simulated size $L=32$. 
We plot the connected susceptibilities $\chi_I$ (filled symbols)
and $\chi_R$ (empty symbols)
for three different pairs of the parameters $I$ and
  $j_0^\prime$. On each panel we plot the curves for six values of the
  island spacing $l$. The peaks of $\chi_I$ (which mark the $T_1$
  transition) depend only weakly on $\ell$. In all panels
  $j_0^\prime=0.072$ }
\label{fig:D2}
\end{center}
\end{figure}

The situation is far clearer for two-dimensional islands (see
Fig.~\ref{fig:D2}), where we still have a finite temperature
thermodynamic transition for isolated islands in the limit of large
sizes. For mesoscopic islands, the crossover between unordered and
ordered island depends more weakly on island size than in the linear
chains case.  Our numerical simulations show that the temperature
$T_1$ does not depend on the island spacing, or the dependence is very
weak. This effect has been also reported in experimental results on
non-mesoscopic island samples~\cite{RESNICK}.  Also the dependence of
$T_1$ on island size is very weak.

\begin{figure}[h]
\begin{center}
\includegraphics[width=0.95\columnwidth]{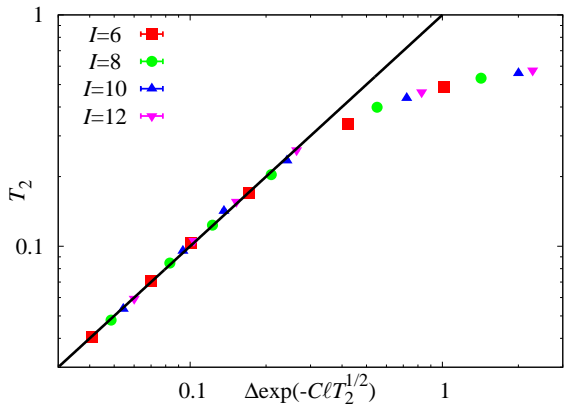}
\caption{A scaling plot of the $T_2$ transition temperature for arrays of D=2
  island based on the behavior $\Delta \exp(-C \ell T_2^{1/2})$ as
  suggested by Eq.~\ref{eq:proximity} ($L=32$ data).}
\label{fig:D2T2}
\end{center}
\end{figure}

We try a more quantitative approach studying the depression of $T_2$
by increasing the inter-island spacing.  We take as an estimate for
$T_2$ the midpoint of the temperature interval bracketing the peak at
its half-height. The dependence of $T_2$ on $\ell$ and $I$ for the largest
simulated array size $L=32$ is
shown in Fig.~\ref{fig:D2T2}.

Following Ref.~\cite{ELEY1}, we notice that
$T_2(\ell)$ compares well to a
proximity-effect prediction
\begin{equation}
  T_2 = \Delta\exp(-C\ell \sqrt{T_2})\;,
\label{fig:T2PROX}
\end{equation} 
corresponding to the solid straight line in Fig.~\ref{fig:D2T2},
suggesting a diverging $\ell(T_2=0)$; we report in
Table~\ref{tab:results} our best fit estimates of the parameters
$\Delta$ and $C$.

\begin{table}[b]
\begin{center}
\begin{tabular*}{0.7\columnwidth}{@{\extracolsep{\fill} }ccc}
\hline
$I$  & $\Delta$ & $C$\\
\hline
6  & 3.74(10) & 0.935(6)\\
8  & 5.21(15) & 0.890(6)\\
10 & 7.56(28) & 0.889(7)\\
12 & 8.15(26) & 0.840(6)\\
\hline
\end{tabular*}
\caption{Best fit estimates of $\Delta$, $C$ parameters in
  Eq.~\ref{eq:proximity} from $T_2$ data for various island size $I$
  ($D=2$, $L=32$, $\ell=8,12,16,24$). Data points for the shortest
  distances $\ell=2,4$ have been excluded from the fits. The
  chi-square per degree-of-freedom values vary between $2.4$ and $3.7$ and
  the quality-of-fit parameters between $0.11$ and $0.32$. Uncertainties
  are gnuplot estimates corrected as in~\cite{YOUNG}.  }
\label{tab:results}
\end{center}
\end{table}

\begin{figure}[h]
\begin{center}
\includegraphics[width=0.95\columnwidth]{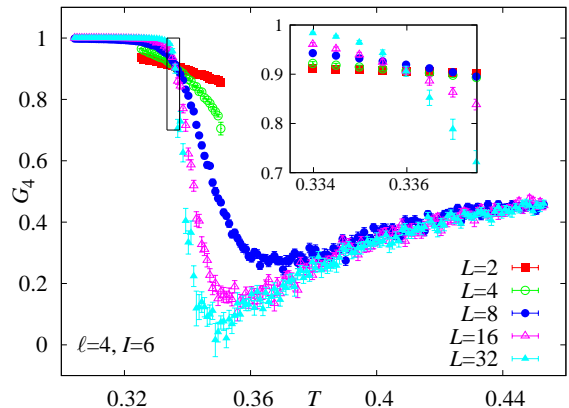}
\caption{Binder cumulant $G_4$ versus $T$ for
$D=2$-islands with $I=6$ and $\ell=4$ and for the five simulated sizes ($L$). In
  the inset we show in detail the region near $T\sim0.335$, where
  different Binder curves cross.}
\label{fig:binder}
\end{center}
\end{figure}

We have run more accurate numerical simulations in the temperature
region close to the $T_2$ transitions, with a four times smaller
cooling rate and ten times more measurements.  We measured the Binder
cumulant
\begin{equation}
  G_4=\frac{1}{2}\left(3-\frac{\thav{\left(\vec{M}^2\right)^2}}
  {\thav{\vec{M}^2}^2}\right)\, .
\label{eq:binder}
\end{equation}
The value of $G_4$ at the $T_2$ transition point is
universal~\cite{AMIT}; we report data for $I=6$, $\ell=4$ and various
array sizes $L$ in Fig.~\ref{fig:binder}. Note that the dips in the
curves of $G_4$ for the largest system sizes in Fig.~\ref{fig:binder}
are due to the breakdown of the $O(2)$ internal symmetry introduced by
the anisotropic form of $\tens{J}^\prime$ in Eq. (\ref{eq:jprime}).
This is the behavior one would expect,
since at high temperature the $G_4$ value
depends on the symmetry of the system.
The fluctuations of the magnetization in the
infinite volume limit are Gaussian distributed and the Binder
parameter for the two-component magnetization of a XY system should
approach the value $G_4(T\rightarrow \infty)=0.5$, whereas for Ising
spins the corresponding high-temperature value is $G_4(T\rightarrow
\infty)=0$. When long range order in the system builds up at low
temperatures, the value of the Binder parameter must approach unity:
$G_4(T\rightarrow 0)=1$.  The data in Fig.~\ref{fig:binder} show that
$G_4$ is not monotonically increasing when the temperature decreases:
it starts at a value around $0.5$ but, as soon as the inter-island
term becomes more important with respect to the intra-island
interaction in the Hamiltonian, the effects of the Ising symmetry sets
in and in proximity of the critical region, just above $T_2$, the
value of $G_4$ drops to low values, as expected for an Ising system.
The minimum of the dip decreases as the system size $L$ increases.

\begin{figure}[h]
\begin{center}
\includegraphics[width=0.95\columnwidth]{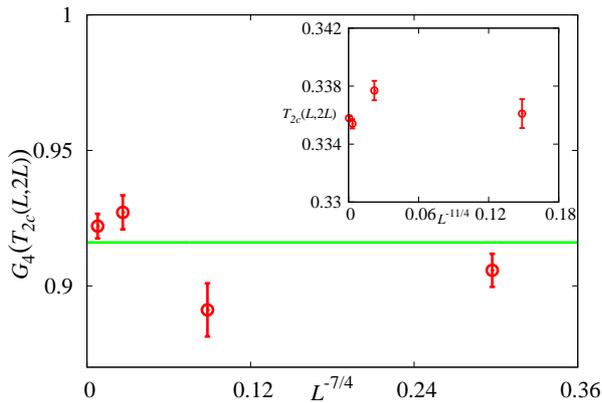}
\caption{Binder cumulant $G_4(T_{2c}(L,2L))$, for
  $D=2$-islands with $I=6$ and $\ell=4$, computed at the crossing
  point of the Binder curves of lattice sizes $L$ and $2 L$, denoted
  as $T_{2c}(L,2L)$, as a function of $L^{-7/4}$. We have marked, with
  a green horizontal line, the $G_4$ value for the two dimensional
  Ising model: $G_4=0.9160386$.  In the inset we show the behavior of
  the temperature-value of the crossing point $T_{2c}(L,2L)$ of the
  Binder curves as a function of $L^{-11/4}$. We have assumed the
  scaling of the two dimensional Ising model: $T_{2c}(L,2L)-T_2 \sim
  L^{-\Delta-1/\nu}$ and $G_4(T_{2c}(L,2L)) - G_4(T_2) \sim
  L^{-\Delta}$. We have used the exact value $\nu=1$ and the
  conjectured one for the correction-to-scaling exponent
  $\Delta=7/4$. For more details, see Ref. \cite{SOKAL}.}
\label{fig:binder_crossing}
\end{center}
\end{figure}

Moreover, the critical value of the Binder cumulant (which is
universal) for the two-dimensional Ising model is known to great
accuracy~\cite{SOKAL}. The values $G_4=0.9160386(24)$ compares
extremely well with our value of the Binder parameter at crossing,
close to $T_2\sim 0.335$ (for comparison, from the position and width
at half-height of the peak of the susceptibility for the same
simulated system we obtain $T_2\simeq 0.340\pm0.004$) We report in
Fig.~\ref{fig:binder_crossing} the details of the crossing of the
Binder curves.  This provides a clear numerical evidence for a second
order phase transition in the two-dimensional Ising universality
class.  The asymptotic value of the crossing points of the Binder
cumulants curves (see inset of Fig. ~\ref{fig:binder_crossing}),
$T_{2c}(L,2L)$ which asymptotically tends to $T_2$, is clearly
different from zero. We finally remark that the value of the Binder
cumulant below the critical temperature is almost unity as expected
asymptotically (as the size of the system goes to infinity).

\section{Conclusions}

It has been very difficult to resolve the crossover at $T_1$ (internal
island ordering) and the $T_2$ transition (inter-island ordering) in
the case of one-dimensional islands. In $D=1$ and in the limit of
large islands sizes we expect $T_1\rightarrow0$ and no thermodynamic
transition at finite temperatures. In the range of simulated island
sizes the measured crossover temperature at which the mesoscopic
island order internally is as small as $T_2$. The island internal
magnetization remains small and no clear maximum of the
susceptibilities signals a crossover down to $T_2$.  At $T_2$ the
inter-island interaction couples the fluctuations of the
magnetizations of neighboring islands, making them coherent: at this
point spins inside each islands starts to align to the average field
of neighboring islands. The mesoscopic character of the islands is
crucial: in the limit of large islands we expect the fluctuations of
local magnetization to be too small (we did not try an experiment in
that direction). Then, although $J^\prime\ll J$, it is the $T_2$
transition that drives both inter-island and internal ordering. As the
inter-island spacing grows the depression of $T_2$ implies the
depression of $T_1$, too. This is compatible with the counterintuitive
requisite that $T_1\rightarrow0$ as $J'\rightarrow0$ discussed above.

Since islands in experimental setups are not laid down on the
substrate as unidimensional chains of columnar grains, the $1D$ system
is not really connected to the experimental setup we analyze here. The
$2D$ island system is, on the contrary, closer to the physical system
we want to understand, and resolving the two transitions is easier in
the case of the $D=2$ system. The $T_1$ transition for $D=2$
macroscopic islands and for $J^\prime\ll J$ is expected to be of the
Kosterlitz-Thouless type.  Since we consider mesoscopic, finite
islands, we observe, as expected, a ``long-range'' ordering of the
islands.  We expect a spin ordering crossover at a temperature value
$T_1$ which does not go to zero with the island size as fast as in the
$D=1$ case.  The model with planar islands is capable of describing
the depression of the global superconductivity transition down to very
low temperature at large lattice spacing.  The transition temperature
we measure by locating the peaks of response functions remains finite
for moderate-to-large inter-island edge-to-edge spacing. $T_2$
approaches a zero value only for very large inter-island spacing. We
cannot detect any dependence of the $T_1$ transition temperature on
$\ell$.  At small $\ell$ values, the value of $T_1$ is influenced by
nearby islands only because of global coherence driving internal
ordering.  We did not try a full finite-size scaling analysis of the
$T_2$ transition, which for well-separated coupling scales $j_0$ and
$j_0^\prime$ is in the Ising universality class.  On the simulated
length scales there are no relevant effects of the mesoscopic
character of the islands on the behavior of $T_2$, which behaves at
large inter-island spacings as expected by the choice of the
dependence of the coupling constants on $\ell$ and $T$.

Appropriate variations of the basic model we have discussed here
could lead to interesting developments in the study of the superconducting
transition in arrays of SNS junctions. We think it is an interesting
starting point to understand many striking experimental evidences, as
for instance the dependence of the transition temperatures on island
thickness, or the strong depression of the $T_1$ and $T_2$ transitions.

\begin{acknowledgement}
We thank Kay Kirkpatrick for introducing us to the model studied in this work 
and her and Jack Weinstein for interesting discussions.  
This work was partially supported by European Union through Grant No. PIRSES-GA-2011-295302,
and ERC Grant No. 247328, by the Ministerio de Ciencia y
Tecnolog\'{\i}a (Spain) through Grant No. FIS2013-42840-P, and by the Junta de
Extremadura (Spain) through Grant No. GRU10158 (partially founded by
FEDER).
\end{acknowledgement}

\end{document}